\documentclass[doublecol]{epl2}

\title{Self-citation corrections for the Hirsch index}
\author{Michael Schreiber
 }
 \institute{
 Institut f\"ur Physik, Technische Universit\"at Chemnitz, 09107
  Chemnitz, Germany }

\abstract{
  I propose to sharpen the index $h$, proposed by Hirsch as a useful
  index to characterize the scientific output of a researcher, by
  excluding the self-citations. Performing a self-experiment and
  also discussing in detail two anonymous data sets, it is shown that
  self-citations can significantly reduce the $h$ index in
  contrast to Hirsch's expectations. This result is confirmed by an
  analysis of 13 further data sets.}

\begin{document}
\maketitle

\section{Introduction}
About one year ago the physicist Hirsch \cite{Hirsch} proposed an
easily computable index $h$ as an estimate of the visibility,
importance, significance, and broad impact of a scientist's
cumulative research contribution. This index $h$ is defined as the
highest number of papers of a scientist that received $h$ or more
citations. Of course, for many people "it is distasteful to reduce
a lifetime's work to a number" \cite{Kel}. For others $h$ is an
"elegantly simple" measure \cite{Cro}, which allows an easy
comparison of the scientific achievement of a scientist in an
unbiased way by a single number. As it can be determined easily by
ordering the publication list according to the number of citations
which is for example possible using the Science Citation Index
provided by Thomson ISI in the Web of Science (WoS) data base, it
has received immediate attention in the public \cite{Bal} and the
physics community \cite{Bat,Pop,Leh} and is already widely
recognized as a convenient measure in evaluations. Already a
significant amount of literature in informetrics
\cite{Eg1,Eg2,Eg3,Gl2,Van,Ros} has been dealing with this measure
of visibility of a scientist. Different data sets have been
evaluated to identify the most highly cited scientists in various
fields \cite{Hirsch,Bat,Cro, Gl3}. A comparative study on
committee peer review \cite{Bor} of post-doctoral researchers in
biomedicine suggested that the Hirsch index is indeed a promising
(rough) measurement of the quality. The statistical correlation of
the Hirsch index with standard bibliometric indicators and peer
judgement was shown to be quite high for 147 chemistry research
groups in the Netherlands \cite{Raan}. A critical analysis of the
Hirsch index of 187 evolutionary biologists and ecologists from
the editorial boards of seven journals illustrates the risk of
indiscriminate use of the index \cite{Kel}. A quantitative
investigation of the statistical reliability \cite{Leh} has cast
doubts on the accuracy and precision of the Hirsch index.
Nevertheless the interest in this measure continues to grow
\cite{Bat,Eg2,Eg1,Eg3,Gl2,Van,Raan,Lia,Eg4,Eg5,Gl1,Ban}.\\
It was shown \cite{Eg3} that the Hirsch index notion can be
extended to the general framework of information production
processes and that any system has a unique Hirsch index. Banks
\cite{Ban} has extended it to an index for scientific topics and
compounds in order to identify hot topics and interesting
materials. The Hirsch-Banks index is defined in analogy to $h$ as
the highest numbers of papers in a particular field or on a
specific compound that received $h$ or more citations. This
extension has also received a lot of attention even beyond the
scientific community, identifying nanotubes, nanowires and quantum
dots as the most interesting topics in recent years. Other
generalizations concern the comparison of entire research groups
by their Hirsch index \cite{Raan} and the utility
for assessing the impact of journals \cite{Bra,Van}.\\
When identifying hot topics, it is obvious that one will be
dealing with a set of publications which are heavily cited within
the field which means that they are probably most often cited by
people working on the same topic, {\it i.e.} by the same set of
people who have written these publications. However, when
assessing the scientific achievement of an individual scientist,
the analogous kind of citations within the data set, namely the
self-citations should ideally not be included, because they are
not reflecting the impact of a publication. Of course,
self-citations increase the $h$ index, but Hirsch has argued that
the effect is relatively small and that the necessary corrections
for $h$ would involve only very few if any papers. An analysis of
a group of scientists in ecology and evolution \cite{Kel},
however, showed an average decrease of 12.3\%. In contrast, the
Hirsch indices of 31 influential scientists in information science
dropped only between zero and three, on average by 0.9, or 6.6\%,
when self-citations were excluded \cite{Cro}. In the present
investigation I demonstrate that the influence of self-citations
on the Hirsch index can be drastic, in particular for younger
scientists with a low Hirsch index. Three different ways to
sharpen the Hirsch index will be proposed.\\

\section{Data base}
It is a rather time-consuming task to identify all self-citations.
Because of self-interest and of the fact that it is relatively
easy to check ones own publications and citations I first
performed a self-experiment and investigated several ways to
determine the self-citations by myself and by my co-authors.
Excluding them, my Hirsch index dropped by 18\%. Then I also
analyzed the publications of a somewhat older colleague who is
working in a more topical field in a mainstream area. In contrast,
I also investigated the records of a somewhat younger colleague,
working in a less attractive field, who has published fewer
papers. Their Hirsch indices also dropped significantly by 13\%
and 46\%, respectively.\\
Before analyzing the self-citations, one has to make sure that the
data base is correct. This concerns the usual difficulties, that
different persons with the same name and same initials are found.
The often suggested solution to check the affiliation is rather
complicated when researchers are concerned who have changed
between various places. Moreover, my own university is an example,
why the correlation with the affiliation is often misleading,
because we not only changed our name between faculty, department
and institute; but also between Hochschule, Technical University,
and University of Technology; and further from Karl-Marx-Stadt via
Chemnitz-Zwickau to Chemnitz. Another problem in establishing the
data base is the possible different way of spelling names, which
is particulary evident for the transliteration of {\it e.g.}
Russian authors, or names which have changed {\it e.g.} by
marriage. In principle, for the identification of the
self-citations the same difficulties occur. However, it is quite
unlikely that a manuscript is cited by a different scientist with
the same name, so that this problem does not occur in practice. On
the other hand, different ways of spelling an author's name or
entirely different names of the same author can easily mask
self-citations so that care should be taken in these cases. Of
course, missing citations because of misspelled names cannot be
avoided, because they do not show up at all in the search. The
data sets used below have been carefully checked with respect to
the mentioned difficulties. In my own case the WoS search yielded
754 results out of which only 268 were my own publications. The
full list would give me a flattering, but wrong $h=46$ instead of
$h^{\rm A}=27$ (The superscript is used to distinguish the
different data sets.). The names of both colleagues whose
publications are analyzed in detail below are not so common, so
that in their cases the analysis was relatively easy, because
nearly all papers which were found in the ISI data base for their
names were really published by these colleagues. For the set B
with 282 papers I analyzed only the 131 publications with 10 or
more citations and found just two which did not appear in this
author's publication lists. For the set C I confirmed that 87 of
the listed 91 papers should be attributed to the colleague. In
both cases there was no influence on the Hirsch index.

\begin{figure}
\onefigure[width=8cm]{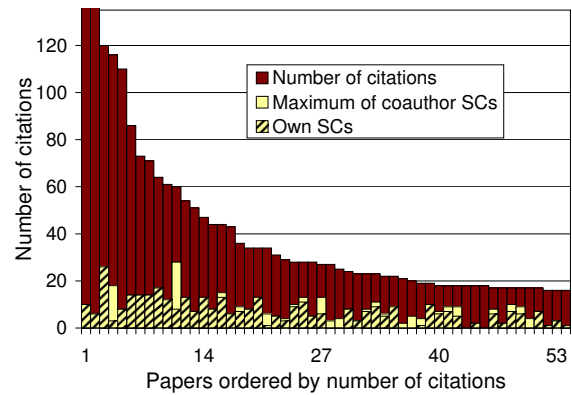}%{Fig1right.eps}
\caption{Number of citations for my 54 most cited papers (dark
grey/brown), own self-citations (hatched), and maximal number of
citations by one of the co-authors including myself (light
grey/yellow).} \label{fig.1}
\end{figure}

\section{Self-citations}
Before coming to the analysis of these data sets, let me comment
on the question, why self-citations may appear. One reason is,
that they are really needed in the manuscript in order to avoid
repetition of previously described experimental setups,
theoretical models, as well as results and conclusions which may
be necessary for the discussion in a certain manuscript but need
not be repeated in this manuscript. Such self-citations are of
course completely legitimate. A second reason for self-citations
is that probably everybody knows his own previous manuscripts best
and therefore it is easier to refer to these own papers when a
citation is required in a given context for a certain argument.
This practice is already questionable, at least when the number of
such self-citations is relatively high. The third reason for
self-citations is certainly disreputable: Due to the
ever-increasing number of evaluations which are based on citation
counts, it is of course tempting to enhance one's citation count
by referring to the own papers for this very purpose. The Hirsch
index is vulnerable to such practice, because it is a single
number which can be relatively easily enhanced by specifically
citing those papers for which the citation count is close to but
below the critical value $h$. For example, in my own case (see
fig.~1) just one citation of my 28th paper would be sufficient to
increase the Hirsch index. However, this paper happens to be first
manuscript that I have ever co-authored so that its "limited
period of popularity" \cite{Red} has long ended, it is also not a
"sleeping beauty" \cite{Red} and therefore it is unlikely to be
cited by somebody else. Therefore I would have to cite it myself,
if I want this paper to have an effect on my Hirsch index. In
future, when the Hirsch index has become -- as I expect -- more
popular, such manipulations might become more severe. In any case,
even the perfectly legitimate self-citations mentioned above
should not be included in any measure of scientific achievement,
therefore the self-citations should be excluded.

\section{SCCs of the first kind: \boldmath {$h_o$}}
The problem is now how to identify self-citations. In the WoS
search one can obtain the names of up to 100 citing authors for a
given paper and how often these people cited the respective paper.
Thus it is easy to identify how often somebody has cited his own
paper. I call this the self-citation corrections (SCCs) of the
first kind. The respective data are shown in figs.~1, 2, and 3 for
the example data sets mentioned above. In my own case, see fig.~1,
eight papers dropped below the critical value of $h^{\rm A}=27$,
five of them even below the value $h_o^{\rm A}=24$ (The subscript
is used to label the different SCCs.). Fortunately two manuscripts
with the full citation count between 24 and 27 remained in that
range even after the SCCs had been taken into account.
Consequently, my Hirsch index was reduced only to $h_o^{\rm
A}=h^{\rm A}-5+2=24$, not to $h^{\rm A}-8=19$. Of course, due to
the strongly fluctuating number of self-citations, the
publications have to be reordered by the number of citations after
the SCCs have been taken into account. The respective result is
shown in fig.~4, confirming $h_o^{\rm A}=24$. For the data set B
in fig.~2, the SCCs are often drastic, like 53 self-citations for
the fifth paper, but usually leaving still a significant number of
other citations. Consequently the SCCs do not influence the Hirsch
index very strongly, they lead to a reduction from $h^{\rm B}=38$
to $h_o^{\rm B}=34$, as shown in fig.~5. In the case C, however,
the SCCs in fig.~3 are so significant, that the citation counts of
all manuscripts fall below the value $h^{\rm C}=13$. However, 7 of
these manuscripts have a corrected count of 7 or more citations,
leading to the new $h_o^{\rm C}=7$. Out of the 12 manuscripts,
which originally had between 7 and 12 citations, two remain in
this range but cannot enhance the $h_o^{\rm C}$ value, as shown in
fig.~6.

\begin{figure}
\onefigure[width=8cm]{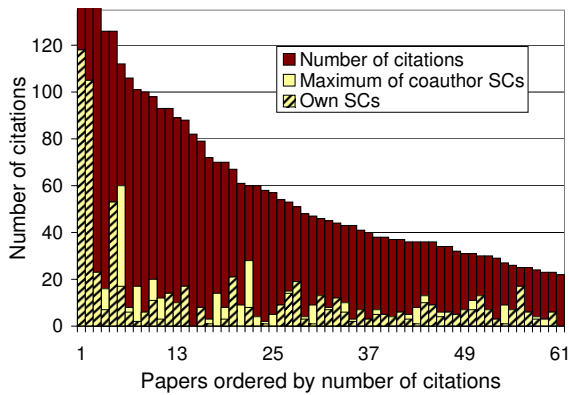}%{Fig2right.eps}
\caption{Same as fig.~1 for the 61 most cited papers in data set
B.}
\end{figure}

\begin{figure}
\onefigure[width=8cm]{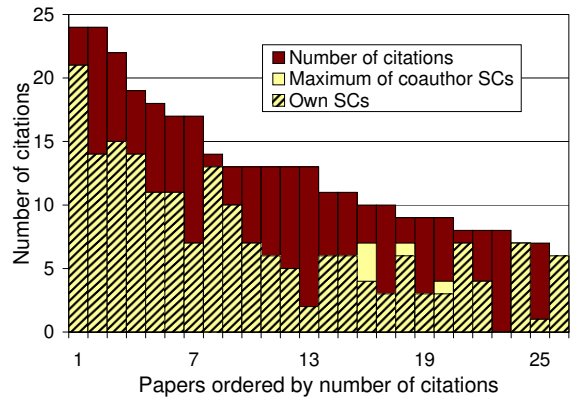}%{Fig3right.eps}
\caption{Same as fig.~1 for the 26 most cited papers in data set
C.}
\end{figure}

\section{SCCs of the second kind: \boldmath {$h_c$}}
Of course, if a paper is cited by one of the co-authors, such a
citation should also not be taken into account. Using again the
above-mentioned ISI list of citing authors for a particular
publication, it is relatively easy to find the co-author with the
highest number of citations for this particular publication. I
call the reduction of the citation count by this number the SCCs
of the second kind. For long author lists, on first sight the
analysis appears to be not so straightforward, because the WoS
summaries show at most 3 authors. However, the "Format for Print
Page" displays all co-authors. In my own case, the number of
citations for several manuscripts dropped significantly more by
the SCCs of the second kind than by those of the first kind, as
can also been deduced from fig.~1, in particular for order numbers
4, 11, 21, 27 -- 29, 36 -- 38, 50. Again a reordering of the
manuscripts had to be performed, the result is included in fig.~4.
The corresponding index $h_c^{\rm A}$, which is corrected for the
(co-)author with the most self-citations, can be determined from
fig.~4 as $h_c^{\rm A}=23$. That means, that the SCCs of the
second kind did further reduce my Hirsch index, but only slightly
although the citation counts of several papers dropped. For the
two colleagues, the respective data are also included in figs.~2,
5 and 3, 6, respectively. In case B, sometimes a co-author was an
even more enthusiastic self-citer, see e.g.~for the sixth paper in
fig.~2, with 60 self-citations. Nevertheless, as this occurred
again mostly for papers with a large citation count, the effect on
the Hirsch index is small, it is reduced to $h_c^{\rm B}=33$. In
the case C rarely a co-author was more enthusiastically citing the
own manuscripts than the investigated author himself, therefore in
this case the Hirsch index remains at the value $h_c^{\rm
C}=h_o^{\rm C}=7$.

\begin{figure}
\onefigure[width=8cm]{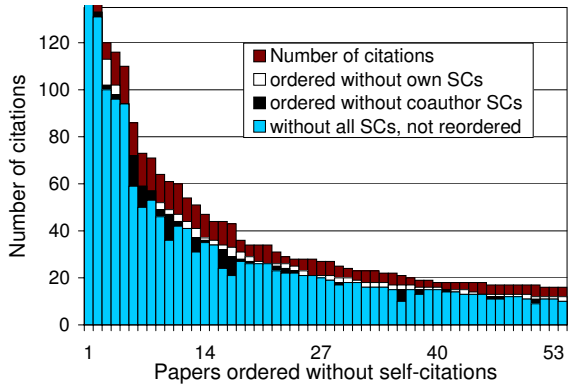} \caption{Number of citations as in
fig.~1 (dark grey/brown), without my own self-citations, reordered
(white), without maximal number of any co-author self-citations,
reordered (black), and without cumulative co-author
self-citations, \underline{not} reordered (medium grey/blue). Note
that the latter histograms conceal the previous ones, so that in
particular the columns of 2nd and 3rd kind often do not show up,
because they are not different from the 3rd and/or 4th kind. The
reordering is not restricted to the 54 papers in fig.~1 but
comprises the full data set.} \label{fig.1}
\end{figure}

Analyzing the author list for the citations of a particular
publication, it is straightforward to identify all co-authors as
long as they appear among the set of 100 citing authors to which
ISI displays are limited. Of course, the effort is significantly
higher than for the SCCs of the second kind, because now one has
to look for all co-author names in often long lists of citing
authors; usually one has to check the complete lists, because some
co-authors, {\it e.g.} typically PhD students never appear.
Therefore I have performed this analysis completely only for my
own publications and for the relatively small data set in fig.~3.
Summing the self-citations of all co-authors of course overshoots
the aim, because the counted self-citations are not just additive
as two authors of a paper may have written another paper together,
citing the first one, which would be counted as a self-citation
for both co-authors. This overestimate can be so severe that it
can lead to negative values for the citation count of papers which
are heavily cited by several co-authors. Nevertheless I have
analyzed the data in figs.~1 and 3 after subtracting the sum of
all self-citations for each paper, resulting in a lower limit for
the corrected Hirsch index of $h_l^{\rm A}=20$ and $h_l^{\rm
C}=5$, respectively. For the data set B, the same analysis was
performed only for the publications with 30 or more citations and
yielded $h_l^{\rm B}=29$. (Note that this result confirms that it
is sufficient to analyze the publications with more than 29
citations.)

\begin{figure}
\onefigure[width=8cm]{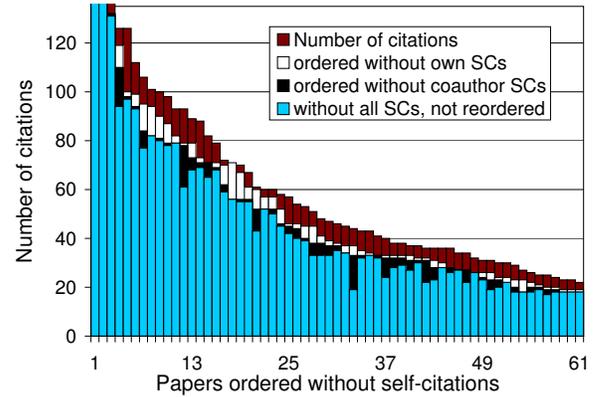} \caption{Same as fig.~4, for data
set B.}
\end{figure}

\begin{figure}
\onefigure[width=8cm]{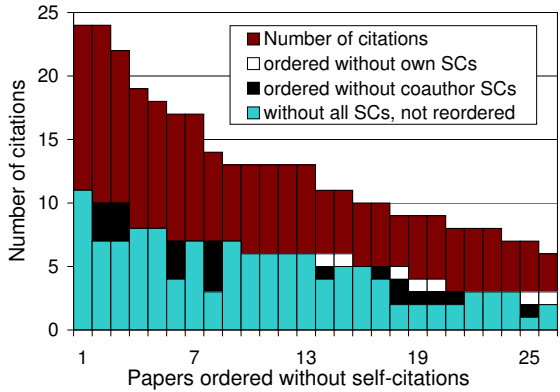} \caption{Same as fig.~4, for data
set C.}
\end{figure}

\section{SCCs of the third kind: \boldmath {$h_s$}}
The correct way of taking multiple co-author self-citations into
account is obviously to check every citing paper for
co-authorship. This yields the SCCs of the third kind. That
requires an enormous amount of tedious work, which can be done
relatively easy for one's own publications although it is still
quite time consuming and error prone. Fortunately one can do this
in the ISI citing author list by checking (ticking off) all
co-author names and viewing the data, which gives a list and thus
the number of cumulative self-citations of all co-authors. The
results are included in figs.~4 -- 6, and the analysis yields a
reduction of the Hirsch index to the sharpened Hirsch index
$h_s^{\rm A}=22$, $h_s^{\rm B}=33$, and $h_s^{\rm C}=7$,
respectively. It can be seen that the effect on the number of
citations for many publications is zero compared to the SCCs of
the second kind. Accordingly the reduction of the Hirsch index
from $h_c$ to $h_s$ is small or zero. Therefore it is a rather
safe assumption, that it is usually sufficient to perform this
analysis only for those papers which are ranked in the vicinity of
the index $h_c$ defined
above taking the SCCs of the second kind into account.\\
For the data set B of fig.~2 which is based on a large number of
publications with a large number of citations and, most important
for the amount of correlations, usually many co-authors, an
analysis of the SCCs of the third kind for only the 22
publications with a citation count between 28 and 56, i.e.,
between 85\% and 170\% of $h_c^{\rm B}$ appeared to be appropriate
a priori. In retrospect, it would have been more than sufficient
to determine the cumulative self-citations for the 10 papers with
a citation count between $h_c^{\rm B}$ and about $1.2~h_c^{\rm
B}$, finding that although 3 out of these publications dropped
below the value of $h_c^{\rm B}=33$, the remaining were just
sufficient to keep $h_s^B$ at $h_c^{\rm B}$. In fact, in this
particular case, even an analysis of the 4 papers with a citation
count of exactly $h_c^{\rm B}$ would have been enough. On the
other hand, starting from the full citation counts (i.e. not
taking first the SCCs of the second kind into account) one would
have had to analyze at least the citations of 26 papers falling
originally into the range between $0.85~h^{\rm B}$ and
$1.6~h^{\rm B}$, in order to reach the correct $h_s^{\rm B}=33$.\\
For my own case an analysis of the 15 publications between
$0.85~h_c^{\rm A}$ and $1.7~h_c^{\rm A}$ yields the correct
$h_s^{\rm A}=22$, but a restriction to the 6 papers in the range
between $h_c^{\rm A}$ and $1.2~h_c^{\rm A}$ already misses one
(the 17th in fig.~4) out of the 3 whose citation counts drop below
$h_c^{\rm A}$. Starting from the full citation counts, the range
of $0.85~h^{\rm A}$ to $1.7~h^{\rm A}$ comprising 21 papers would
have been just sufficient to determine $h_s^{\rm A}$
correctly.\\
For case C, the range $0.85~h_c^{\rm C}$ to $1.7~h_c^{\rm C}$
covers 13 publications including the ones most cited (after
excluding the SCCs of the second kind). Therefore it is not
surprising that this range is more than sufficient to determine
$h_s^{\rm C}$. In fact, also in this case an analysis of the 4
papers at $h_c^{\rm C}$ would have been enough to corroborate the
value $h_s^{\rm C}=7$, although 2 of these drop below $h_s^{\rm
C}$. On the other hand, starting from the original citation counts
(i.e. without considering SCCs), even the range of $0.85~h^{\rm
C}$ to $1.7~h^{\rm C}$ would have been insufficient for a correct
analysis, one would have to start as low as $0.6~h^{\rm C}$ and
include also the most cited papers to obtain the correct value
of $h_s^{\rm C}$.\\
In table 1 the discussed values are compiled. The relative
reduction from $h$ to the sharpened Hirsch index $h_s$ is
considerable. Interestingly, the absolute decrease is nearly the
same in all the above analyzed cases, namely 5 or 6, although very
different publication and citation patterns distinguish the cases.
It is therefore only a conjecture, when I infer that such an
absolute reduction might by
typical.\\
In order to test this conjecture I have analyzed also a fourth
data set D reflecting the achievements of a prominent scientist in
La Jolla, again finding an absolute decrease of 5 which, however,
amounts to a reduction of only 10\%, because Hirsch's index is
rather high. I have then investigated another 12 data sets of
physicists which I know rather well so that it has been possible
with a reasonable amount of effort to make sure that the data base
is correct, in particular excluding publications of different
persons with the same name and same initials, but on the other
hand including publications with deviating spellings of the name
(mainly due to missing second initials or an umlaut in the name).
The obtained results are also included in table 1 as data sets
E--P, sorted by the (original) Hitsch index. It turned out, that
the absolute decrease from the original Hirsch index to the
sharpened Hirsch index was 3 or 4 in most cases, thus being
somewhat smaller than in the first 4 data sets. However, this is
still significant, especially noting that the relative reduction
is between 20 and 25\% in most cases. Of course, due to the small
values, a difference of 1 or 2 in the results should not be
overvalued. However, as one example I note that the sharpened
Hirsch index makes a distinction between data sets C and H much
clearer than the original Hirsch index. On the other hand
comparing data sets C and O, one finds from the original index the
reasonable assumption that C is better than O. But the sharpened
index suggests the opposite order. Data sets J and K are
distinguishable by the sharpened index, but not by the original
index.

\begin{table}[htb]
\caption{Hirsch index without and with SCCs (data in sets A--D
compiled August 2006, in sets E--P January 2007). The total number
of publications, the highest citation count, and the relative
reduction of the Hirsch index to the sharpened index are also given
for each data set.} \label{tab:1}
\renewcommand{\arraystretch}{1.1}
\begin{tabular}{crrrrrrr}
%{llp{1cm}llp{1cm}ll}
  \hline
  \vspace{0cm}
  &&&\multicolumn{4}{c}{kind of SCCs}&  \vspace{0.1cm}\\
  \cline{4-7}
  &&&none&1st &2nd &3rd &
  \vspace{0.1cm} \\
  \cline{1-8}
   data&total&max.&\multicolumn{4}{c}{Hirsch index}&
   \vspace{0.1cm}\\
  \cline{4-8}
%{set}&\raisebox{0.2cm}[-0.2cm]{no.}
  \raisebox{0.2cm}[-0.2cm]{set}&\raisebox{0.2cm}[-0.2cm]{no.}
  &\hspace{-0.1cm}\raisebox{0.2cm}[-0.2cm]{count}&$h$&$h_o$&$h_c$&$h_s$
  &$1-\frac {h_s}{h}$\\
  \hline
 A&268&178&27&24&23&22&18\%\\
  B&280&420&38&34&33&33&13\%\\
 C&86&24&13&7&7&7&46\%\\
  D&183&468&50&47&46&45&10\%\\
 E&322&73&20&17&16&16&20\%\\
 F&63&279&19&16&15&15&21\%\\
   G&66&149&15&14&13&12&20\%\\
 H&51&112&15&14&13&12&20\%\\
 I&72&55&14&11&11&10&29\%\\
  J&77&47&13&10&10&9&31\%\\
   K&47&108&13&13&11&11&15\%\\
 L&61&40&12&10&9&9&25\%\\
 M&46&53&12&11&10&10&17\%\\
  N&60&79&10&7&7&7&30\%\\
 O&44&41&10&9&8&8&20\%\\
  P&15&25&5&4&4&3&40\%\\
  \hline
  \end{tabular}
\end{table}

\section{Conclusion and outlook}
In conclusion, Hirsch's conjecture that usually only very few if
any papers need to be dropped from the $h$ count, if
self-citations were taken into account, has been shown to be
unrealistic. It may well be true for the prominent physicists that
he has mentioned in his paper \cite{Hirsch}. Roediger \cite{Roe}
has also argued about the self-citations that for "people with
very high counts, they aren't much of a problem." However, for the
average scientist, this is not valid. I even believe it to be a
good guess that for younger scientists with comparatively low
Hirsch index, the influence of the SCCs is often relatively
strong. Most of the data sets in table 1 are from younger
scientists. But it is at this stage of the career where the Hirsch
index is or will be probably most often used for the assessment of
the scientific achievements of a scientist, be it for a promotion
or for the comparison with competitors for an open position. One
might argue from table 1 that the Hirsch index "only" renormalizes
by about 20\% due to SCCs and therefore remains to be a useful
measure even with SCCs. For more prominent people this may be
true, but for younger scientists the discussed deviations from the
average reduction are important. Consequently, the Hirsch index
should be used with reasonable care, and it would be good policy
to take the SCCs into account. As mentioned above, it is
straightforward and easy to determine the SCCs of the first kind
and it is also relatively easy to calculate the corrections of the
second kind. Taking the third kind, i.e., the cumulative
self-citations into account is of course the method of choice, but
it is rather difficult to execute, unless an automatic correlation
between author lists and citing author lists can be performed.\\
Other corrections may also be reasonable in particular when
comparing people working in different areas. It has already been
observed by Hirsch \cite{Hirsch} that citation patterns in
different fields vary significantly. This was quantified
\cite{Igl} in terms of a scaling factor. Another correction with
the (average) number of co-authors has been proposed
\cite{Hirsch,Bat,Ba2,Roe} and a large impact especially in physics
was found \cite{Bat,Ba2}. As the Hirsch index usually increases
with the number of publications, it has been suggested to compare
it with the average $h$ for scientists in the same field and the
same number of publications in order to detect those researchers
who "clearly deviate from world standards" \cite{Igl}. One should
also be aware that the general search in the WoS data base does
not take into account books, book chapters, or conference
proceedings. For some fields these are less relevant, while in
other fields they might be decisive for the impact of a
scientist's research. Of course, it would also be interesting to
investigate, how an individual's Hirsch index increases with time
\cite{Hirsch,Lia}.\\
Based on a large data set of publications \cite{Red1} the
distribution of citations has been studied and a growing random
network model was used to describe the citation statistics
\cite{Kra,Kra1}. Citation patterns in a more homogeneous community
in high energy physics have also been analyzed and modelled in
detail \cite{Leh2,Leh1}. As already mentioned, when one wants to
identify hot fields of research the citations within a certain
community are of interest and should be measured, so that the
self-citations might even have some value and need not necessarily
be excluded. On the other hand, it is well known that there exists
schools, sometimes also called citation cartels, whose members try
to increase their visibility by citing mostly friends and family.
It would be an interesting exercise to exclude citations within
such a school from the determination of the Hirsch index. This can
in principle be done by compiling a list of all co-authors with
whom a certain scientist has published any paper and to exclude
from the citation list of every manuscript every citation by
anybody from this list. When this cumulative co-author list
increases with time because new co-authors appear on the list,
then the self-citation-corrected count of older manuscripts can be
decreased and thus the index can also decrease with time, which is
not possible due the SCCs discussed above, nor it is
possible for the original proposal of Hirsch.\\
In any case, I believe that at least the own citations, i.e. the
self-citations of the first kind, which can be most easily
determined, should be excluded from any evaluation, because they
can be most easily manipulated by the author. The temptation to
increase one's Hirsch index oneself should be avoided, even though
some journals explicitly suggest to their authors to cite
themselves or other papers of the journal in order to increase the
impact factor. This is of course understandable from their
business point of view, but it is questionable from the scientific
point of view.\\


\begin{thebibliography}{10}

\bibitem{Hirsch}
\Name{Hirsch~J.E.} \REVIEW {Proc. Natl. Acad.
Sci. U.S.A.}{102}{2005}{16569}.

\bibitem{Kel}
\Name{Kelly~C.D. \and Jennions~M.D.} \REVIEW {Trends in Ecology and
Evolution}{21}{2006}{167}.

\bibitem{Cro}
\Name{Cronin~B. \and Meho~L.} \REVIEW {J. Am. Soc. Inf. Sci.
Techn.}{57}{2006}{1275}.

\bibitem{Bal}
\Name{Ball~P.} \REVIEW {Nature}{436}{2005}{900}.

\bibitem{Pop}
\Name{Popov~S.B.} {A parameter to quantify dynamics of a
researcher's scientific activity}, {\it arXiv:physics}/0508113.

\bibitem{Leh}
\Name{Lehmann~S., Jackson~A.D., \and Lautrup~B.} {Measures and
mismeasures of scientific quality} \REVIEW {\it arXiv:physics
\rm/0512238 and\it Nature} {444} {2006} {1003}.

\bibitem{Bat}
\Name{Batista~P.D., Campiteli~M.G., Kinouchi~O., \and
Martinez~A.S.} \REVIEW {Scientometrics}{68}{2006}{179}.

\bibitem{Eg2}
\Name{Egghe~L.} \REVIEW {ISSI Newsletter}{2}{2006}{8}.

\bibitem{Eg1}
\Name{Egghe~L.} {\it J. Am. Soc. Inf. Sci. Techn.}, to appear
{(2006)}{}.

\bibitem{Eg3}
\Name{Egghe~L. \and Rousseau~R.} \REVIEW {\it
Scientometrics}{69}{2006}{121}.

\bibitem{Gl2}
\Name{Gl\"anzel~W.} \REVIEW {Scientometrics}{67}{2006}{315}.

\bibitem{Van}
\Name{Vanclay~J.K.} \REVIEW {Scientist}{20}{2006}{14}.

\bibitem{Ros}
\Name{Rousseau~R.} {Simple models and the corresponding h- and
g-index}, {\it http://eprints.rclis.org/archive}/00006153.

\bibitem{Gl3}
\Name{Gl\"anzel~W. \and Persson~O.} \REVIEW {\it ISSI Newsletter}
{1} {2005} {15}

\bibitem{Bor}
\Name{Bornmann~L. \and Daniel~H.D.} \REVIEW
{Scientometrics}{65}{2005}{391}.

\bibitem{Raan}
\Name{van~Raan~A.F.J.} \REVIEW {Scientometrics}{67}{2006}{491}.

\bibitem{Lia}
\Name{Liang~L.} \REVIEW {\it Scientometrics}{69}{2006}{153}.

\bibitem{Eg4}
\Name{Egghe~L.} \REVIEW {Scientist}{20}{2006}{14}.

\bibitem{Eg5}
\Name{Egghe~L.} \REVIEW {\it Scientometrics}{69}{2006}{131}.

\bibitem{Gl1}
\Name{Gl\"anzel~W.} \REVIEW {Science Focus}{1}{2006}{10}.

\bibitem{Ban}
\Name{Banks~M.G.} \REVIEW {\it Scientometrics} {69} {2006} {161}.

\bibitem{Bra}
\Name{Braun~T., Gl\"anzel~W., \and Schubert~A.} \REVIEW
{Scientist}{19}{2005}{8}.

\bibitem{Red}
\Name{Redner~S.} \REVIEW {Physics Today}{58}{2005}{49}.

\bibitem{Roe}
\Name{Roediger~H.L.} \REVIEW {APS Observer} {19} {2006} {4}

\bibitem{Igl}
\Name{Iglesias~J.E. \and Pecharroman~C.} {Scaling the $h$-index
for different scientific ISI fields}, {\it arXiv:physics}/0607224.

\bibitem{Ba2}
\Name{Batista~P.D., Campiteli~M.G., Kinouchi~O., \and
Martinez~A.S.} {Universal behaviour of a research productivity
index}, {\it arXiv:physics}/0509048.

\bibitem{Red1}
\Name{Redner~S.} \REVIEW {\it Eur. Phys. J. B} {4} {1998} {131}.

\bibitem{Kra}
\Name{Krapivsky~P.L., \and Redner~S.} \REVIEW {\it Phys. Rev. E}
{63} {2001} {066123}.

\bibitem{Kra1}
\Name{Krapivsky~P.L., Redner~S., \and Leyvraz~F.} \REVIEW {\it
Phys. Rev. Lett.} {85} {2000} {4629}.

\bibitem{Leh1}
\Name{Lehmann~S., Jackson~A.D., \and Lautrup~B.} \REVIEW {\it
Europhys. Lett.} {69} {2005} {298}.

\bibitem{Leh2}
\Name{Lehmann~S., Lautrup~B., \and Jackson~A.D.} \REVIEW {\it
Phys. Rev. E} {68} {2003} {026113}.







\end{thebibliography}
\end{document}